\documentclass[fleqn,usenatbib]{mnras}
\usepackage{newtxtext,newtxmath}
\usepackage[T1]{fontenc}
\DeclareRobustCommand{\VAN}[3]{#2}
\let\VANthebibliography\thebibliography
\def\thebibliography{\DeclareRobustCommand{\VAN}[3]{##3}\VANthebibliography}

\usepackage{graphicx,subfigure}	
\DeclareMathOperator{\Var}{\widehat{Var}}
\usepackage{ulem} 

\usepackage{hyperref}

\usepackage{xcolor}

\title[DM-power: high precision DM for FRBs]{DM-power: an algorithm for high precision dispersion measure with application to fast radio bursts}

\author[Lin et al.]{Hsiu-Hsien Lin$^{1,2}\thanks{E-mail:\href{hsiuhsil@alumni.cmu.edu}{hsiuhsil@alumni.cmu.edu}}$,
Robert Main$^3$,
Ue-Li Pen$^{1,2,4,5,6}$,
Robert Wharton$^{7}$,
Marlon Luis	Bause$^{3}$,
\newauthor
Suryarao Bethapudi$^{3}$,
Dongzi Li$^{8}$,
Fang Xi	Lin$^{2,9}$, 
Visweshwar Ram Marthi$^{10}$,
Laura G	Spitler$^{3}$.
\\
$^{1}$Institute of Astronomy and Astrophysics, Academia Sinica, Astronomy-Mathematics Building, No. 1, Sec. 4, Roosevelt Road, Taipei 10617, Taiwan \\
$^{2}$Canadian Institute for Theoretical Astrophysics, 60 St. George Street, Toronto, ON M5S 3H8, Canada\\
$^{3}$Max-Planck-Institut f{\"u}r Radioastronomie, Auf dem H{\"u}gel 69, D-53121 Bonn, Germany \\
$^{4}$Canadian Institute for Advanced Research, 180 Dundas St West, Toronto, ON M5G 1Z8, Canada \\
$^{5}$Dunlap Institute for Astronomy and Astrophysics, University of Toronto, 50 St George Street, Toronto, ON M5S 3H4, Canada \\
$^{6}$Perimeter Institute of Theoretical Physics, 31 Caroline Street North, Waterloo, ON N2L 2Y5, Canada \\
$^{7}$NASA Postdoctoral Program Fellow, Jet Propulsion Laboratory, California Institute of Technology, Pasadena, CA 91109, USA\\
$^{8}$Cahill Center for Astronomy and Astrophysics, MC 249-17 California Institute of Technology, Pasadena CA 91125, USA \\
$^{9}$Department of Physics, University of Toronto, 60 St. George Street, Toronto, ON M5S 1A7, Canada \\
$^{10}$National Centre for Radio Astrophysics, Tata Institute of Fundamental Research, Post Bag 3, Ganeshkhind, Pune - 411 007, India} 

\date{Accepted XXX. Received YYY; in original form ZZZ}

\pubyear{2015}

\begin{document}
\label{firstpage}
\pagerange{\pageref{firstpage}--\pageref{lastpage}}
\maketitle

\begin{abstract}
We present DM-power, a new method for precisely determining the dispersion measure (DM) of radio bursts, and apply it to the Fast Radio Burst (FRB) source FRB~20180916B.  Motivated by the complex structure on multiple time scales seen in FRBs, DM-power optimizes the DM by combining measurements at multiple Fourier frequencies in the power spectrum of the burst.  By optimally weighting the measurements at each Fourier frequency, DM-power finds a burst DM that effectively incorporates information on many different burst timescales.  We validate this technique on simulated Gaussian pulse profiles with a precision down to $\sigma_{\rm DM} \sim 0.001~{\rm pc~cm}^{-3}$, and then apply it to bursts from pulsar B0329+54 and FRB~20180916B.  The precision of these DM measurements are sufficient to measure a statistically significant variation in DM over a $\approx 2$~hr span.  While this variation could be the result of electron density variations along the line of sight, it is more like that the observed variation is the result of intrinsic frequency-dependent burst structure that can mimic a dispersive delay.
\end{abstract}


\section{Introduction}\label{section: Introduction}
Fast Radio Bursts (FRBs) are energetic (peak flux $\sim$Jy), ms-duration, radio transients from cosmological origin with unknown physical mechanism \citep{2007Sci...318..777L, 2019A&ARv..27....4P, 2019ARA&A..57..417C}. In the past decade, over six hundred FRBs have been reported \citep{2016PASA...33...45P, 2021ApJS..257...59C}. Among them, about fifty FRBs have been observed to repeat \citep{2016Natur.531..202S, 2019Natur.566..235C, 2019ApJ...885L..24C, 2020ApJ...891L...6F, 2023arXiv230108762T}. Two repeating FRBs, FRB 121102 and FRB 20180916B, show periodic activity \citep{2020MNRAS.495.3551R, 2021MNRAS.500..448C, 2020Natur.582..351C}. 

The propagation of an FRB through the tenuous electron plasma between the source and observer imparts a frequency dependent delay to the observed signal. The dispersive delay between two observing frequencies is: 
\begin{equation}
\Delta t = D \times \left(\nu_{\rm lo}^{-2}-\nu_{\rm hi}^{-2}\right)\times \mathrm{DM} \label{equation: delta_t}
\end{equation}
where $\Delta t$ is the time delay in seconds, $D = 1/(2.41\times 10^{-4})~{\rm MHz}^{2}~{\rm cm}^{3}~{\rm pc}^{-1}~{\rm s}$ is the dispersion measure constant \citep{2020arXiv200702886K}, $\nu_\mathrm{lo}$ and $\nu_\mathrm{hi}$ are the low and high frequencies in MHz, respectively, and DM is the dispersion measure defined as 
\begin{equation}
\mathrm{DM} = \int_{0}^{d}n_{\rm e} \, dl\label{equation: dm_define}
\end{equation}
with an units of ${\rm pc~cm}^{-3}$ \citep{2019A&ARv..27....4P, 2019ARA&A..57..417C}. If the DM of a source is known, the frequency-dependent delays can be removed and the radio signal can be aligned across frequencies in the dynamic spectrum (i.e. the waterfall plot).  However, high precision DM measurements may be confounded by other frequency dependent effects like intrinsic burst drifting,  other propagation effects, and burst morphology evolution \citep{2019ApJ...876L..23H, 2021ApJ...923....1P}.

Over 50 models have been proposed to explain the origin of FRBs \citep{2019PhR...821....1P}, including bursts from magnetars and super-giant pulses from pulsars \citep{2017ARA&A..55..261K, 2019MNRAS.490L..12B}. Possible explanations for repeating FRBs include precessing neutron stars \citep{2021ApJ...909L..25L, 2020ApJ...892L..15Z, 2020ApJ...895L..30L} and binary systems \citep{2020ApJ...893L..39L, 2020ApJ...893L..26I}. The various models predict different properties. For example, the linear polarization-angle swing may related to the scenario of magnetar or supernova remnant pulsar \citep{2016MNRAS.458L..19C}, and the linear polarization-angle swing may depend on the periodic phase \citep{2021ApJ...909L..25L}, or the range of DM variation may indicate various physical origin of the repeating FRBs \citep{2017ApJ...847...22Y, 2020ApJ...893L..39L}. Previously, the DM variation in the order of $\sim$10$^{-3}$ pc cm$^{-3}$ has been observed in PSR B1957+20 \citep{2018Natur.557..522M}, a black widow pulsar with a companion, across the orbital phase near the eclipse.

Since repeating FRBs generally show complex morphology in the dynamic spectrum \citep{2019ApJ...876L..23H}, the traditional way that measuring the DM through maximizing the peak signal-to-noise ratio (S/N) would not align the drifting substructure of the intensity across frequencies \citep{2019ApJ...876L..23H}. Instead, several algorithms have been proposed to improve the DM measurement \citep{2019ApJ...876L..23H, 2019ascl.soft10004S, 2020MNRAS.499L..16M, 2021MNRAS.505.3041P, 2022MNRAS.509.2209M}, and the typical uncertainty of the DM is in the order of $\sim$0.1 pc cm$^{-3}$ \citep{2019ApJ...885L..24C, 2020ApJ...891L...6F}. 

Measuring burst DMs at high precision could apply for searching for periodic variations in FRB DMs, which may indicate a binary motion. \citet{2020ApJ...893L..31Y} mention that the 16-day periodicity of FRB~20180916B can change and the intrinsic pulse structure could vary on minutes timescales. Hence, high-precision DM measurements could be used to probe the the local environment around pulsars, magnetars, and repeating FRBs \citep{2017ApJ...847...22Y}. 

\citet{2019ascl.soft10004S} created the DM-phase\footnote{\url{https://github.com/danielemichilli/DM_phase}} algorithm to precisely measure the DM of FRBs with complex structure. The DM-phase algorithm works by maximizing the structure of the waterfall by taking the time derivative of the frequency integrated pulse profile at different DMs. 

On the other hand, \citet{2020MNRAS.499L..16M, 2022MNRAS.509.2209M} measured the DM and the uncertainty in the power spectrum by fitting a Gaussian profile to the noise-free profile at the highest Doppler frequency in the power spectrum (i.e. the highest noise-free profile), where the Doppler frequency is defined as the Fourier transform of time. The peak value and the full width at half maximum (FWHM) represents to the optimized DM and the DM uncertainty, respectively. However, the highest noise-free frequency mode is hard to be determined for fainter bursts, and the uncertainty given by the FWHM is overestimated.

In this paper, we present a new algorithm (hereafter: DM-power\footnote{\url{https://github.com/hsiuhsil/DM-power}}) that yields very precise DM measurements with uncertainties as little as $\delta {\rm DM} \sim 0.001~{\rm pc~cm}^{-3}$ for the simulated Gaussian pulse profiles mentioned in the paper, which is up to 10 times more precise than other methods. However, the ultimate limit to the precision for the sources considered in the paper will depend on the burst morphology, the S/N, the timing resolution, and other effects that we will discuss in the paper. Our algorithm is similar to previous microstructure-optimized algorithms such as DM-phase, but with several important differences. First, instead of averaging over all of the Fourier frequencies in the power spectrum and measuring the DM in the averaged profile, we optimize the sub-structure in each of the rebinned Fourier frequencies, which is less sensitive to the time gridding and therefore the measurement is more robust. Second, as we do not know the probability distribution function (PDF) of the power spectrum in the Fourier frequencies and the DM steps, we apply a non-parametric method, the bootstrap test, to determine the DM and its uncertainty. We have validated the algorithm with the simulated Gaussian pulse profiles, and we apply the algorithm to single pulses from PSR B0329+54 find that the DM measurement is more sensitive to a burst with more complex morphology, which is consistent with optimizing the structure in each of the Fourier frequencies. 

In Section \ref{sec: observation}, we briefly review the observation of FRB 20180916B described in \citet{2020MNRAS.499L..16M} as well as the observation of PSR B0329+54 \citep{1968Natur.219..574C}. In Section \ref{sec: Analysis}, we show the measurement of the DM and the uncertainty with the mode-independent method and the bootstrap tests. In Section \ref{sec: discussion}, we compare with DM-phase, show the DM measurement of 56 single pulses from PSR B0329+54, and we discuss the DM  measurement of 11 bursts, including the noise effects, the morphology effect, and the degeneracy. Finally, we summarize the results in Section \ref{sec: summary}.

\section{Observations and Data}\label{sec: observation}

\subsection{FRB 20180916B}\label{sub:FRB 20180916B}
\citet{2020MNRAS.499L..16M} reported 15 bursts of FRB 20180916B detected by the upgraded Giant Metrewave Radio Telescope (uGMRT) on 2020 March 24 (MJD 58932) and 2020 June 30 (MJD 59030). We select 11 bright bursts, in which the peak flux is higher than 0.22 Jy, as well as further manually masking radio frequency interference (RFI) for the analysis in this paper. Figure \ref{figure: waterfalls_before_after} shows the 11 bursts with the initial DM of 348.82 pc cm$^{-3}$ and the corresponding optimized DM from this work. Figure \ref{figure: waterfalls_res} (See Appendix. \ref{sec:The dynamic spectrum before and after the DM optimization}) shows the differences between the original and the optimized case. 

The data properties are described in \citet{2020MNRAS.499L..16M}. In brief, the data was recorded by the coherently dedispersed (CD) phased array beam with 2048 frequency channels at 550-750~MHz, a timing resolution of 327.68~$\mu$s, and coherently dedispersed to 348.82~pc~cm$^{-3}$. 

The method in this paper addresses the issue of inter-channel DM variation. In the CD phased array mode, any structure of the intra-channel less than the timing resolution (i.e. 327.68 $\mu$s) is dedispersed to the default DM value (i.e. 348.82 pc cm$^{-3}$). The coherent dedispersion setup records data at the inputted DM, while the incoherent dedispersion records data at DM of 0 pc cm$^{-3}$. In either case, we miss the micro-structure within the frequency channel. In Section \ref{sec: Analysis}, we will show the DM measurement is not constrained by the intra-channel resolution. 

\subsection{PSR B0329+54}\label{subsec: 56 pulses from PSR B0329+54}
We observed PSR B0329+54, a pulsar with a period of $\sim$0.715 seconds with bright single pulses of broadband, multi-component, and mode-switching features \citep{2005AJ....129.1993M, 2011ApJ...741...48C, 2019MNRAS.484.2725B}. We used the CD phased array beam of the uGMRT for a total duration of 285 seconds on 2020 February 21 (MJD 58900). The data has a total of 2048 frequency channels at 550-750 MHz (same as for the 15 bursts of FRB20180916b) with a timing resolution of 81.92 $\mu$s (4 times finer than the observations of FRB 20180916B in \citet{2020MNRAS.499L..16M}). We select a total of continuous 56 single pulses from the observation, which we initially incoherently dedisperse the pulses with a DM of 26.7641 pc cm$^{-3}$ for the analysis in this work.

\subsection{Generating simulated Gaussian pulse profiles}\label{sub: Generating simulated Gaussian pulse profiles}

We simulate Gaussian pulse profiles for the following analysis by using
\begin{equation}\label{equation: gau_sim}
\mathrm{I_{gauss}(f,t)} = \mathrm{A(f)\exp{-\frac{(t-t_{0})}{2w^{2}}}+N(f,t)},  
\end{equation}
where $\mathrm{{I_{gauss}(f,t)}}$ represents the simulated Gaussian pulse profile with the same frequency and time resolution of PSR B0329+54 dataset (see Section \ref{subsec: 56 pulses from PSR B0329+54}) at DM of 0 pc cm$^{-3}$, $\mathrm{A(f)}$ is a constant amplitude across the frequency channels, $\mathrm{t}$ represents the variable in time axis, $\mathrm{t_{0}}$ is the t parameter at 0 second as the center of the Gaussian peak, $\mathrm{w}$ is a parameter to control the width of the pulse profile, and $\mathrm{N(f,t)}$ is a normal distributed noise background. Appendix Figure \ref{figure: gaussian_sim_pulses} shows the 9 simulated Gaussian pulse profiles, which have the range of the peak S/N from 200 to 15 and the width from 33.2 to 1.7 ms.

\section{Analysis}\label{sec: Analysis}

\begin{figure*}
\centering
\includegraphics[width=0.85\textwidth]{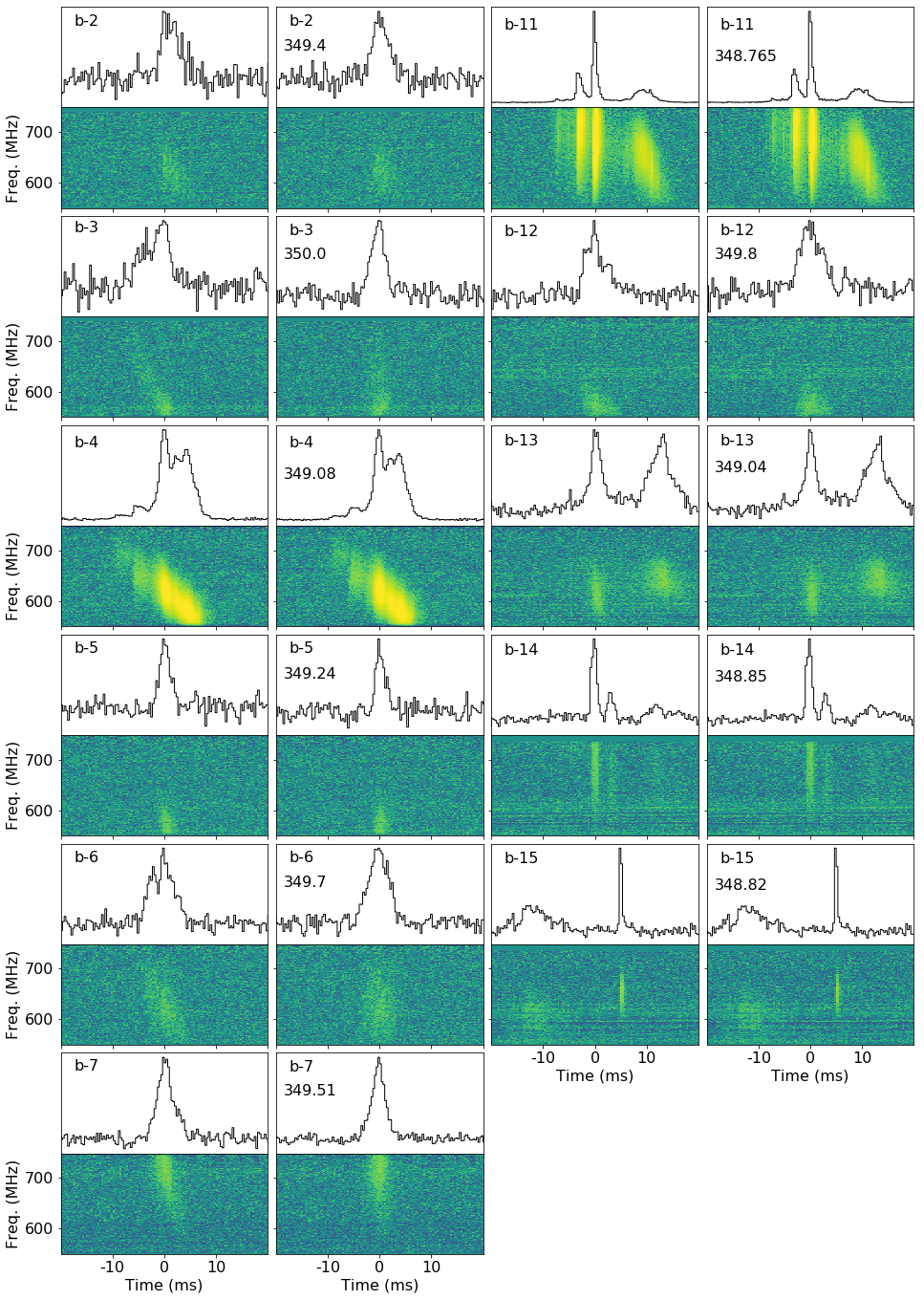}
\caption{\textbf{The waterfall of the 11 bursts selected from \citet{2020MNRAS.499L..16M} with the DM before and after the optimization technique.} The first and third columns show the waterfall at the original DM at 348.82 pc cm$^{-3}$, and the second and fourth columns show the waterfall at correspondingly optimized DM with the value listed, which the measurement is discussed in this paper. For each panel, the top and bottom represents the time-series and dynamic spectrum, respectively. We mark the burst number corresponding to \citet{2020MNRAS.499L..16M} in the time-series.}
\label{figure: waterfalls_before_after}
\end{figure*}

In this Section, we will discuss the DM-power method and demonstrate it on the simulated Gaussian pulse profiles, which was mentioned in Section \ref{sub: Generating simulated Gaussian pulse profiles}. Broadly speaking, the DM-power method takes as input a dynamic spectrum (or waterfall plot) of a burst de-dispersed at an approximately correct DM.  The burst is then decomposed into Fourier modes (or Doppler frequencies) along the time axis, which are independently fitted for a DM. Finally, we combine the DM values from the Fourier modes into a single measurement using an optimal weighting scheme. We measure the DM uncertainty by taking the standard deviation of a sample of measured DM, which we bootstrap the original dynamic spectrum and measure the corresponding DM.

\subsection{Burst Structure at Different Timescales}
As shown in Figure~\ref{figure: waterfalls_before_after}, FRBs show structure on a range of different timescales, some of which may be more sensitive to changes in DM than others.  To fully utilize this information, we want to decompose each burst into different modes at different timescales and then measure DMs for each mode.  We can illustrate this with Fourier decomposition.

For each burst, we have intensity data, $\mathrm{I}\left(\mathrm{t};\nu_\mathrm{RF}\right)$, as a function of time $t$ and radio frequency $\nu_\mathrm{RF}$.  
We can represent this as a superposition of multiple components along the time dimension:
\begin{equation}\label{equation: intensity}
\mathrm{I}\left(\mathrm{t};\nu_\mathrm{RF}\right) =\sum_k {\mathrm{I}_{k}\left(\mathrm{t};\nu_\mathrm{RF}\right)},
\end{equation}
where, by definition, 
\begin{equation}\label{equation: intensity_decomposition}
\mathrm{I}_{k}\left(\mathrm{t};\nu_\mathrm{RF}\right) =\sum^{\mathrm{N}-1}_{0} {\mathrm{I}\left(\omega_{k}[\mathrm{n}-\frac{\mathrm{N}}{2}]\frac{1}{\mathrm{T_{s}}};\nu_\mathrm{RF}\right)};
\end{equation}
$\omega_{k}$ is the angular frequency of the $k$ component, $\mathrm{T_{s}}$ is the sample time, there are N time samples, n indexes the samples. Each I$_{k}$ therefore represents the decomposed waterfall that corresponds to a specific Doppler frequency (i.e., Fourier frequency) bin, which is obtained as 
\begin{equation}\label{equation: intensity_reconstruct}
\mathrm{I}_{k} = \mathfrak{F}^{-1}[\mathfrak{F}_{k}\mathrm{I}\left(\mathrm{t};\nu_\mathrm{RF}\right)],
\end{equation}
where $\mathfrak{F}^{-1}$ and $\mathfrak{F}$ represents the inverse discrete Fourier Transform with real output (IRFFT) and the discrete Fourier Transform with real output (RFFT) along the time axis, respectively.

The result of this decomposition is a set of time-frequency arrays $\mathrm{I}_{k}(t; \nu_{\rm RF})$ that give the original burst $I(t; \nu_\mathrm{RF})$ filtered at each Fourier frequency.  Thus, we have decomposed the original burst into all of its component timescales. 
The shorter timescales, and hence the higher Fourier frequency components, are more sensitive to the sub-structure of the burst with a more precise DM estimation. Thus, the higher Fourier frequency components are more weighted. This would result in that the algorithm maximising the sharpness of bursts. However, for the sake of better sensitivity and reduced computational costs, it is convenient for us to combine neighboring $\mathrm{I}_{k}(t; \nu_{\rm RF})$ using logarithmic binning of the Fourier modes, which corresponds to the sub-structure in different time-scales. In this case, we can now write Equation \ref{equation: intensity_reconstruct} as
\begin{equation}\label{equation: intensity_reconstruct_log}
\mathrm{I}_{k} = \mathfrak{F}^{-1}[\mathfrak{F}\mathrm{I}\left(\mathrm{t};\nu_\mathrm{RF}\right)]|_{k},
\end{equation}
where the "|$_{k}$" jointly represents binning and indexing the k$^\mathrm{th}$ binned component.

\begin{figure*}
\centering
\includegraphics[width=0.9\textwidth]{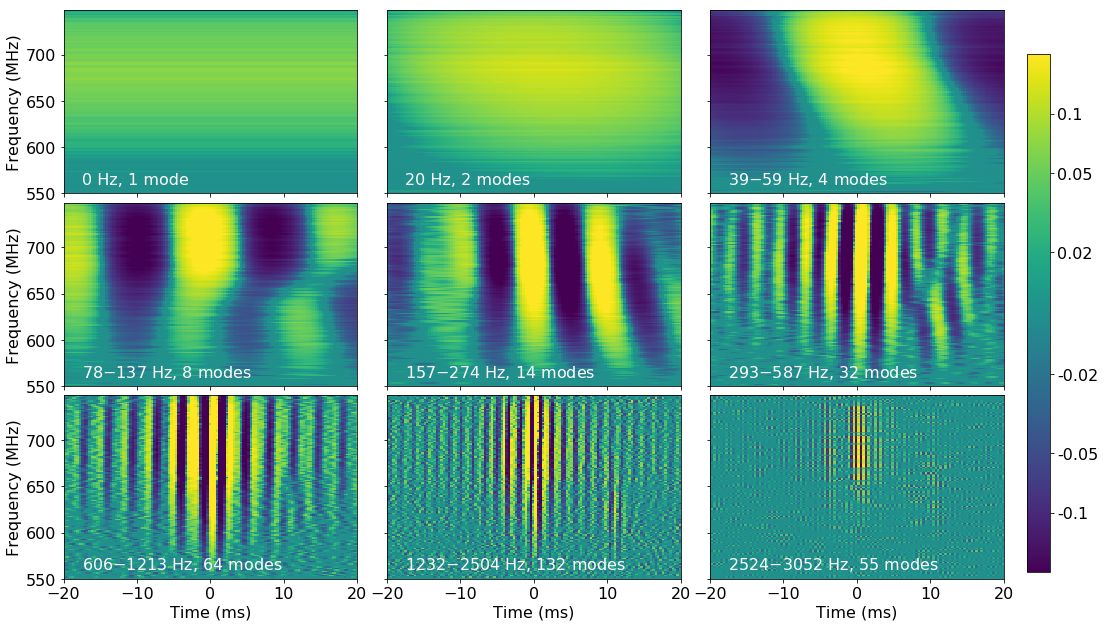}
\caption{\textbf{The IRFFT-reconstructed waterfalls of the b-11 at the original DM of 348.82 pc cm$^{-3}$ across the RFFT frequencies.} Each panel shows the IRFFT-reconstructed waterfall in the corresponding RFFT frequencies as well as the total number of the summed IRFFT modes. The micro-structure becomes evident at high frequency modes, justifying applying different weightings to different Fourier modes for the optimization of DM, which we will describe in Section \ref{subsec: Measure the FFT-mode independent DM and error}. The sum of the 9 IRFFT-reconstructed waterfalls is equal to the original waterfall before the RFFT iterations as shown in the panel in Figure \ref{figure: waterfalls_before_after}.}
\label{figure: log_rfft_waterfall}
\end{figure*}

As an example, we will consider burst b-11 from Figure~\ref{figure: waterfalls_before_after}. Figure~\ref{figure: log_rfft_waterfall} shows the decomposition of b-11 into nine logarithmically spaced Fourier frequency bins. Each panel is effectively a ``Doppler-bandpass'' filtered component of the original burst waterfall, meaning that it shows the waterfall over a range of Doppler frequencies (and corresponding time scales). We start with a 102.24 ms (i.e. 312 time-bins) subset of time-frequency data of burst b-11 de-dispersed at the initial DM of $348.82~{\rm pc~cm}^{-3}$. Following the above steps, we then decompose the burst into 312 Fourier frequencies. We then combine arrays from the 312 Fourier frequencies into 9 logarithmically spaced Doppler frequencies (i.e. Equation \ref{equation: intensity_reconstruct_log}). Figure~\ref{figure: log_rfft_waterfall} shows the 9 waterfalls plots of I$_{k}$ with the corresponding Fourier frequency ranges and the total number of summed Fourier modes. The sum of these 9 waterfalls is equal to the original waterfall.

We have illustrated the decomposition in the time domain here because it is a more intuitive way to show the burst structure on different time scales. However, in practice, we will calculate the DMs exclusively in the Fourier domain as described in the next section.

\subsection{Power Spectrum DM Search}
\label{subsec: Generate the power spectrum} 

In a standard SNR-maximizing DM search, the DM is found by searching  over a range of trial DMs and calculating the resulting SNR of the de-dispersed burst in the time domain.  The trial DM that produces the highest SNR burst is then taken to be the true DM.  Since we want to be sensitive to burst structure on many time scales, we will instead optimize the DM by measuring the power in the Fourier domain.

We start with the dynamic spectrum of a burst that has been de-dispersed to an approximately correct DM by other means (e.g., SNR-maximizing method). For the data analyzed here, we take 102.24~ms worth of data containing the burst (on-pulse) and the same duration not containing the burst (off-pulse), as this is a sufficient duration for searching for DM offsets from -2 to $+2~{\rm pc~cm}^{-3}$. Using a step-size of $0.08~{\rm pc~cm}^{-3}$, we then loop over the trial DMs and apply the incoherent de-dispersion to the dynamic spectrum with a reference frequency at the top of the band, producing a de-dispersed dynamic spectrum for each trial DM. Instead of summing across frequency and calculating the power spectrum at this stage, we first utilize a Singular Value Decomposition (SVD) step to reduce noise.

The Singular Value Decomposition (SVD) is a general technique to optimize the measurement by decomposing the signal into different eigenvalues and eigenfunctions. The higher eigenvalues and corresponding eigenfunctions represent the more dominated features in the signal.
Usually using the leading first one or two modes could reconstruct the leading feature of the system, and the number of leading modes to be used depends on the signal-to-noise of the feature. Hence, the noise background can be suppressed by using the leading eigenvalues and eigenfunctions to reconstruct the noise-free signal \citep{2018MNRAS.475.1323L, 2021MNRAS.508.1947T}. Thus, we apply the SVD to each de-dispersed dynamic spectrum. For the on- and off-pulse regions at each trial DM step, we apply the SVD to the dynamic spectrum to decompose it into the eigenvalues and eigenfunctions in time and frequency 
\begin{equation}\label{equation: svd}
P_{ft} =\sum_n {U_{fn}}{S_{n}}{V_{nt}^{\top}},
\end{equation}
where $P_{ft}$ represents the dynamic spectrum in frequency ($f$) and time ($t$), $n$ is eigenmode number, $S_{n}$ is the eigenvalue, $U_{ft}$ is the eigenfunction in frequency, and $V_{nt}^{\top}$ is the transposed eigenfunction in time. We use the first mode of the eigenfunction in frequency as a weighting function ($W = P_{ft} \times U_{f0}$) to take the weighted mean over frequency channels to produce a de-dispersed weighted profile.  We then take a Fourier transform and make power spectrum for both the on- and off-pulse regions. The off-pulse power spectrum is subtracted from the on-pulse power spectrum, and the result is re-binned into nine Doppler frequency (f$_{D}$) bins with logarithmic spacing.

\begin{figure*}
\centering
\includegraphics[width=0.9\textwidth]{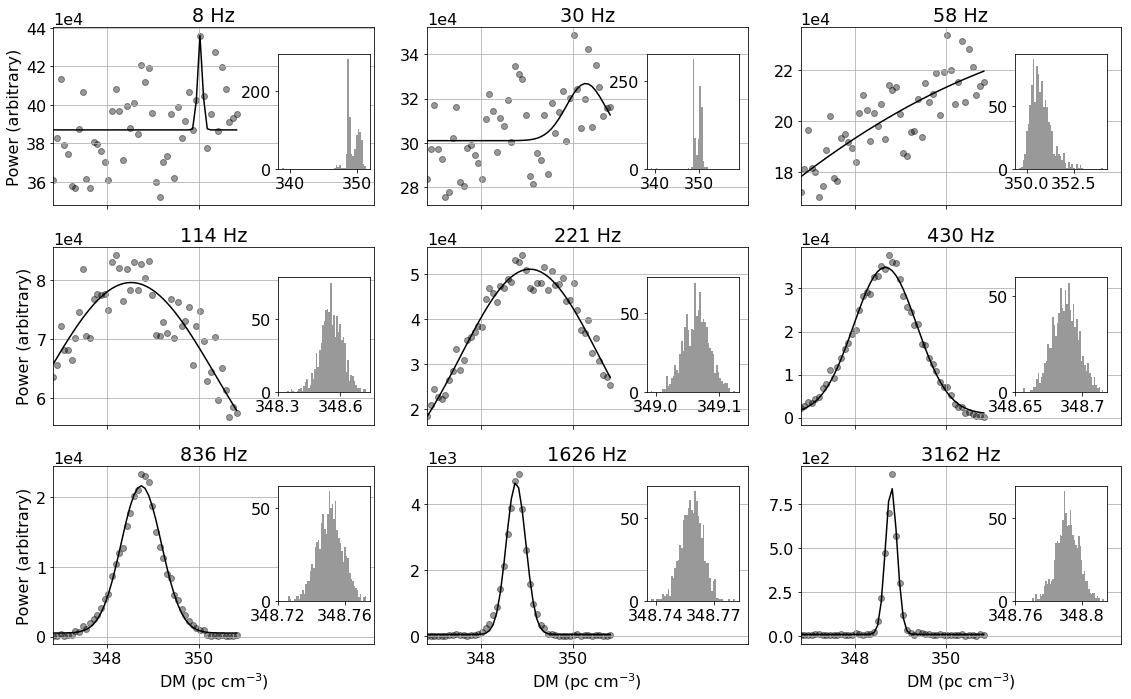}
\caption{\textbf{The power spectrum and the bootstrap tests of b-11.} In each panel, the dots and curve represent for the power spectrum at the corresponding log-rebinning Doppler frequency and the Gaussian fitting curve, respectively. The inserted panel represents the histogram distribution of the peak value from the Gaussian fitting curve by using the bootstrap tests.}
\label{figure: log_power_spectrum}
\end{figure*}

The process described here is repeated for every trial DM, which produces a power value for every trial DM in each of the Doppler frequency bins.  Figure~\ref{figure: log_power_spectrum} shows the result for all nine of the Doppler frequencies. 

\subsection{Measuring the DM in Each Frequency Bin}
\label{subsec: The bootstrap tests}

Using the power values calculated at every trial DM, we can now find the peak DM in each of the Doppler frequency bins by fitting Gaussians as shown in Figure~\ref{figure: log_power_spectrum}. Since the distribution of the power spectrum across Doppler frequencies is unknown, we probe its properties with a bootstrap test. The bootstrap test is a nonparametric approach to estimate the sample distribution in statistics \citep[e.g.][]{1984MNRAS.210P..19B}. In brief, one draws a sample with replacement many times \citep{efron1979bootstrap}, so the distribution of the sample can be statistically understood.

\begin{figure*}
\centering     
\subfigure[b-11]{\label{fig:dm_opt_b11}\includegraphics[width=0.45\textwidth]{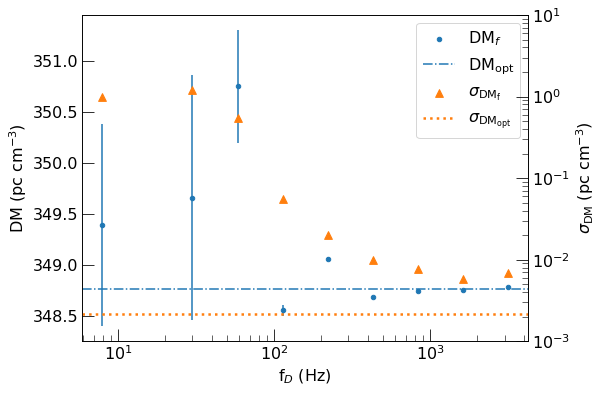}}
\subfigure[b-4]{\label{fig:dm_opt_b4}\includegraphics[width=0.45\textwidth]{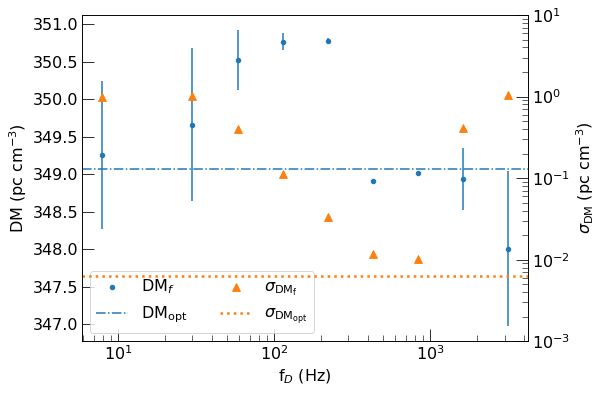}}
\caption{\textbf{The mean and the standard deviation of the fitted DM across log-Doppler frequencies.} At each of the log-Doppler frequencies, the blue circle and the orange triangle represent the mean and the standard deviation of the fitted DM, respectively. The blue dash-dot and the orange dotted line represents the optimized DM and the corresponding uncertainty mentioned in Section \ref{subsec: Measure the FFT-mode independent DM and error}.}
\label{figure: dm_opt_results}
\end{figure*}

The fitting procedure with bootstrapping goes as follows. For each DM step we de-disperse the dynamic spectrum and then create a new dynamic spectrum of the same size by randomly drawing 2048 frequency channels with replacement. We then produce a power spectrum from this new dynamic spectrum and measure the power values in each Doppler frequency bin as described in Section~\ref{subsec: Generate the power spectrum}.  This process is repeated 1000 times for each trial DM to produce 1000 instances of the measured power as a function of trial DM for each Doppler frequency bin.  For each of the Doppler frequencies, we use least squares fitting to fit a Gaussian to each of the 1000 instances of measured power over trial DM. We take the median value of the 1000 fitted Gaussian widths as a fixed value for each Doppler frequency and repeat the fitting to find the DM value where the Gaussian peak occurs.  The distributions of the peak DM values are shown in the inset panels of Figure \ref{figure: log_power_spectrum}.  We take the mean and standard deviation of the peak DM values to be the measured DM and DM uncertainty in each Doppler frequency bin. The results of this fitting process for burst b-11 are shown in Figure~\ref{fig:dm_opt_b11}.

Note that the RFI channels could be duplicated selected during the bootstrap steps, and hence the RFI channels could be correlated. This may results in an underestimated DM uncertainty, as we will discuss in Section \ref{subsec: Measure the FFT-mode independent DM and error}. Hence, we encourage further study on the interplay between DM measurement and RFI contamination.

\subsection{Combined DM Measurement}
\label{subsec: Measure the FFT-mode independent DM and error}
\subsubsection{the optimized DM}
Now that we have measured the DM values and uncertainties for each of the Doppler frequency bins, we want to combine them in an optimal way to get a single DM and DM uncertainty for a burst. To do this, we will adopt the strategy of \citet{1986PASP...98..609H} and weight the optimal DM as: 
\begin{align}
\mathrm{w_{m}} &= \frac{1}{\Var\left(\mathrm{DM_{m}}\right)} = \frac{1}{\left(\sigma_\mathrm{DM_{m}}\right)^{2}}\label{equation: dm_weighting},\\
\mathrm{DM} &= \frac{\sum \mathrm{DM_{m}w_{m}}}{\mathrm{\sum_{m}w_{m}}}\label{equation: dm_total}, 
\end{align}
where $\mathrm{DM_{m}}$ and $\sigma_\mathrm{DM_{m}}$ are the DM value 
and uncertainty measured in the m-th Doppler bin.

\subsubsection{the DM uncertainty}
To determine the uncertainty of the DM measurement, we generate 100 dynamic spectra by applying the bootstrap to the same intensity 100 times, and measure the standard deviation of their DM measurements in Equation \ref{equation: dm_total}. This is so-called Method1.

\begin{figure*}
\centering
\includegraphics[width=0.9\textwidth]{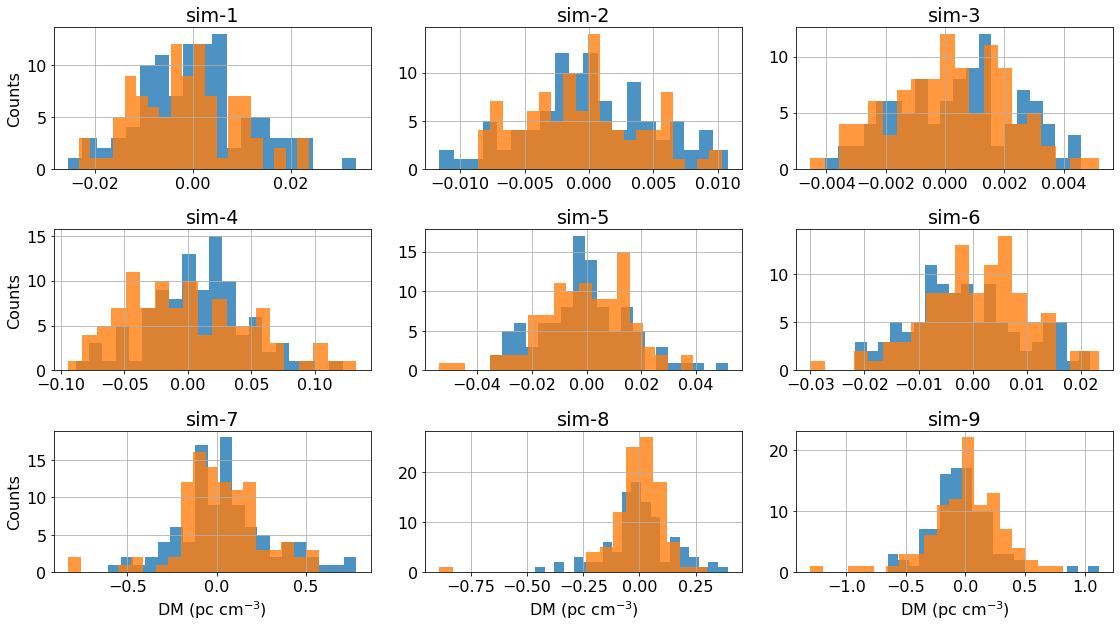}
\caption{\textbf{The histogram of the DM measurements.} Each panel represent the DM measurement for the 9 Gaussian simulated pulse profiles that mentioned in Section \ref{sub: Generating simulated Gaussian pulse profiles}. The blue and orange histogram represents the Method1 and Method2 that mentioned in Section \ref{subsec: Measure the FFT-mode independent DM and error}, respectively.}
\label{figure: Two_methods_DM_uncertainty}
\end{figure*}

The error estimation of the DM-power algorithm is based on the bootstrap technique, which is equilibrium to the normal distribution. Hence, we use the simulated Gaussian pulse profile to validate the approach. For each of the 9 Gaussian pulse profiles that mentioned in Section \ref{sub: Generating simulated Gaussian pulse profiles}, we randomly simulate 100 pulse profiles with the same Gaussian parameters that mentioned in Equation \ref{equation: gau_sim}, and we apply the algorithm to measure their DM. This is so-called Method2. Figure \ref{figure: Two_methods_DM_uncertainty} shows the histogram of the DM measurements of Method1 and Method2. The corresponding mean and standard deviation values are listed in Table \ref{table: table_DM_comparisons}, which shows that Method1 and Method2 are consistent with each other. We use Method1 as the reported $\mathrm{\sigma_{DM}}$ (i.e. DM uncertainty) of DM-power. The DM precision depends on the S/N ratio as well as the width of the burst. For instance, the DM uncertainty reported by DM-power for a pulse with peak S/N of 200 and width of 1.7 ms and a pulse with peak S/N of 15 and width of 33.2 ms is $\sim$0.001 and $\sim$0.1 pc cm$^{-3}$, respectively.

\begin{table}
\centering
\begin{tabular}{c |c | c |c |c |c}
\hline \hline
sim & S/N$\mathrm{_{peak}}$ & width & DM-power & DM-power & DM-phase  \\
           &          & (ms) & Method1 & Method2 & \\
\hline
1	& 200  & 33.2  & 0.0007(100) & -0.0017(100) & 0.002(80) \\
2	& 200   & 8.3  & 0.0001(50)  & -0.0006(50)  & -0.003(23)\\
3	& 200   & 1.7  & 0.0005(20)  & 0.0002(20)   & -0.012(3)\\
4	& 50  & 33.2  &  0.007(40)   & -0.0012(500) & -0.004(123)\\
5	& 50   & 8.3  &  -0.001(20)  & -0.0005(200) & -0.03(6)\\
6	& 50   & 1.7  &  -0.001(10)  & 0.0008(100)  & -0.006(63)\\
7	& 15  & 33.2  &  0.05(30)    & 0.03(20)     & -0.0008(2700)\\
8	& 15   & 8.3  &  0.01(10)    & -0.002(100)  & 0.08(20)\\
9	& 15   & 1.7  &  0.03(30)    & 0.006(300)   & -0.06(26)\\
\hline
\end{tabular}
\caption{\textbf{The comparison of the DM measurements of the nine Gaussian simulated pulse profiles between DM-power and DM-phase.} 
The first, second, and third column shows the number, the peak S/N, and the width of the simulated Gaussian pulse profile as shown in Appendix Figure \ref{figure: gaussian_sim_pulses}, respectively. The fourth and fifth column shows the DM and the uncertainty (i.e. the standard deviation value from the histogram shown in Figure \ref{figure: Two_methods_DM_uncertainty}) of Method1 and Method2, respectively. The third column shows the DM and the uncertainty reported by the DM-phase package. All values are in the unit of pc cm$^{-3}$.
}
\label{table: table_DM_comparisons}
\end{table}

\section{Discussion}\label{sec: discussion}

\subsection{Comparison with DM-phase}\label{subsec:Comparison with the DM-phase}

In this section, we use the 9 simulated Gaussian pulse profiles, the single pulses of PSR B0329+54, and a bright burst of FRB~20180916B to compare DM-power and DM-phase. We find that the DM-power is more sensitive to the bright pulses with sharp features.

\subsubsection{The simulated Gaussian pulse profiles}\label{subsub: the simulated Gaussian pulse profiles}

We further compare the DM measurement with the DM-phase. We use the DM-phase package to measure the DM and the uncertainty of the 9 simulated Gaussian pulse profiles shown in Figure \ref{figure: gaussian_sim_pulses}, and the reported DM values are listed in Table \ref{table: table_DM_comparisons}. We find that generally the DM uncertainty reported by the DM-power is a few times less than the DM-phase.

\subsubsection{Single pulses of PSR B0329+54}

To test the efficacy of this technique, we apply the DM optimization method to 56 single-pulses from PSR~B0329+54, which we mentioned in Section~\ref{subsec: 56 pulses from PSR B0329+54}. We repeat the procedure discussed in Section \ref{sec: Analysis}, using 163.84~ms for each of the on- and off-pulse regions and 1000 bootstrap steps. Appendix Figures~\ref{figure: b0329_gallery_1} and \ref{figure: b0329_gallery_2} show the waterfall plots with the optimized DM for each pulse.

\begin{figure}
\includegraphics[width=0.45\textwidth]{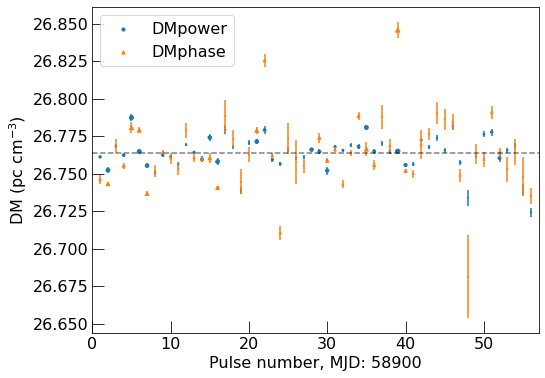}
\caption{\textbf{The DMs, uncertainties, and the peak S/N versus the pulse number of the single pulses from PSR B0329+54.} We apply the DM-power and DM-phase algorithms, which is marked in blue and orange, respectively, to 56 single pulses from PSR B0329+54. Each dot represents the  DM of the 56 bursts, and the dot area is proportional to the peak S/N. The brightest and the dimmest pulses are pulse-39 and pulse-48 with the peak S/N of 3856 and 100, respectively. The error bar corresponds to 68$\%$ confidence interval. The horizontal dashed line represents the initial DM of 26.7641 pc cm$^{-3}$ that reported in \citet{2005AJ....129.1993M}. Note that the DM-phase cannot resolve the DM of pulse-28, and we omit the data point in the plot.}
\label{figure: dm_toa_sn_b0329}
\end{figure}

We also apply the DM-phase to the same dataset. Figure \ref{figure: dm_toa_sn_b0329} shows the DM and the uncertainty, reported by the DM-power and DM-phase\footnote{Note that DM-phase does not work on pulse-28, which probably due to the very sharp feature, and we ignore this data point in Figure \ref{figure: dm_toa_sn_b0329} as well as in the reduced chi-square calculation.} algorithms, and the peak S/N of the 56 single pulses from PSR~B0329+54 in order of the time-of-arrival (ToA). Among the 56 pulses, b-48 has the highest ratio between the DM uncertainty reported by the DM-power and the DM-phase, which the DM uncertainty reported by DM-power is $\sim$7 times more precise than the one reported by DM-phase. The reduced chi-square of the 56 DM values and uncertainties reported by DM-power and DM-phase is 26.2 and 38.2, respectively. The DM variation may be due to the micro-structure \citep{2003MNRAS.344.1187K} and the degeneracy effect, including frequency dependent burst morphology, and other propagation effects that we will discuss in Section \ref{subsub: the DM variation and the degeneracy effect}.

On the other hand, the giant pulses of PSR B0329+54 usually show multi-components and sharp features with different amplitudes, which have been in bursts of repeating FRBs \citep{2019ApJ...876L..23H}. Our comparison between DM-power and DM-phase on the giant pulses of PSR B0329+54 suggests that DM-power has a more precise measurement on the DM and the DM uncertainty.

\citet{2015JInst..10C7002T} reported that the annual variation of DM is consistent with the Doppler motion between the Earth and the Sun. On the other hand, \citet{2019MNRAS.487..394T} mentioned that the DM variations on timescales of days by the Solar wind is $\sim$10$^{-4}$ pc cm$^{-3}$. Hence, the daily changes of DM of 10$^{-4}$ pc cm$^{-3}$ is an upper limit estimate on the DM variation for a minute of data. Moreover, the dot size represents the peak S/N, which we do not see an obvious correlation between the peak S/N and either the DM$\mathrm{_{opt}}$ or the $\sigma_{\mathrm{DM_{opt}}}$. 

\subsection{Applying DM-power to FRB~20180916B}

We apply the DM optimization method described in Section~\ref{sec: Analysis} to eleven of the FRB~20180916B bursts reported by \citet{2020MNRAS.499L..16M} and report the results in Table \ref{table: table_DM}. We have selected only the bursts from \citet{2020MNRAS.499L..16M} with peak flux above 0.22~Jy.  The other four bursts were much fainter, with resulting uncertainties above $\sigma_{\mathrm{DM_{opt}}}\sim 0.1~{\rm pc~cm}^{-3}$, which we consider to be dominated by the noise background rather than the FRB signal as discussed in Section~\ref{subsec: noise amplification}.

\subsubsection{A bright burst of FRB~20180916B}

The FRB~20180916B burst b-11 has a rich morphology on multiple timescales with two bright narrow leading peaks and a fainter broader trailing peak with a drifting down pattern.  Hence, we consider this burst an excellent test for comparing our DM-power algorithm to the widely used DM-phase algorithm \citep{2019ascl.soft10004S}.

\begin{figure}
\includegraphics[width=0.45\textwidth]{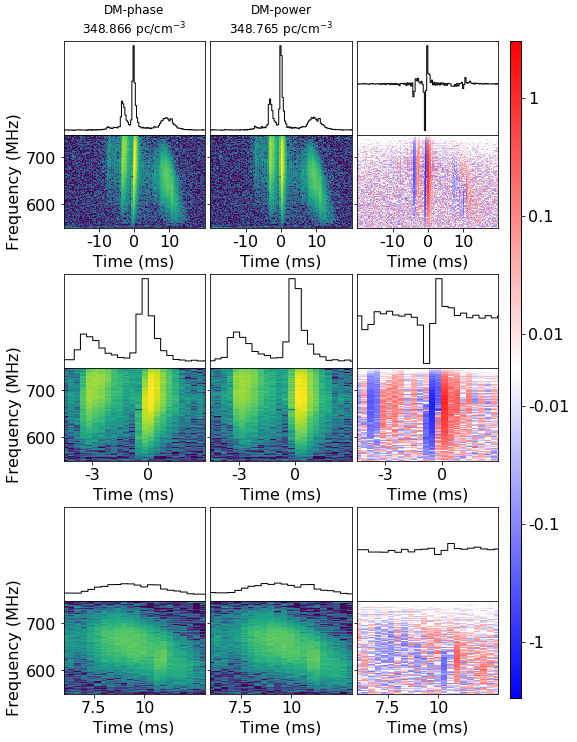}
\caption{\textbf{The comparison between the DM-phase and the DM-power on the b-11.} The first and second column show the waterfall of the b-11 at the optimized DM of the DM-phase (348.866 pc cm$^{-3}$) and the DM-power (348.765 pc cm$^{-3}$), respectively. The third column shows the difference that the second column subtracts from the first column. The second and the third row shows the zoomed version from the first row.}
\label{figure: DM_comparisons}
\end{figure}

For the b-11 burst, DM-phase yields an optimized DM\footnote{We use the default setup, fitting with 10 data points, to get 348.866(3) pc cm$^{-3}$. If we change the data points from 10 to 30 for the fitting, the result is 348.866(10) pc cm$^{-3}$.} of 348.866(3)~pc~cm$^{-3}$, and DM-power gives an optimized DM of 348.765(8)~pc~cm$^{-3}$. Figure~\ref{figure: DM_comparisons} shows the de-dispersed waterfalls for each optimized DM. We notice that the leading-peak at $\sim$0~ms is slightly more dedispersed using the DM from DM-phase.  In contrast, the same peak is straightly aligned using the DM from DM-power. This is consistent with the discussion in Section \ref{section: Introduction} that DM-power is less sensitive to the time gridding, and therefore the DM-power measurement is more robust for a narrower burst with a higher amplitude.

\subsubsection{The DM variation and the degeneracy effect}\label{subsub: the DM variation and the degeneracy effect}

\begin{figure}
\includegraphics[width=0.45\textwidth]{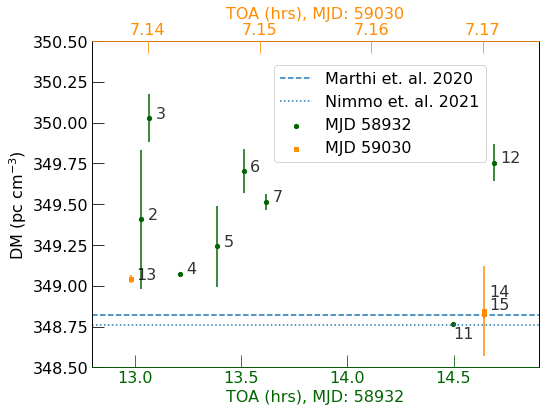}
\caption{\textbf{The optimized DMs versus TOAs.} We apply the optimization technique to eleven bursts in \citet{2020MNRAS.499L..16M}. Each dot represents the optimized DM of the eleven bursts. The error bar corresponds to a 68$\%$ confidence interval. The green and orange dots represent the optimized DM values for the bursts detected on MJDs 58932 and 59030, respectively. The MJDs 58932 and 59030 correspond to the lower and upper horizontal axis, respectively. The dashed line marks DM of 348.82 pc cm$^{-3}$, which is the default value in \citet{2020MNRAS.499L..16M} of MJDs 58932 and 59030 (i.e. in different activity cycles.). The dotted line represents DM of 348.76, which is the optimized DM value reported from the micro-structure analysis by  \citet{2021NatAs...5..594N} } of MJD 58653.
\label{figure: dm_toa}
\end{figure}

The DM-power measured DMs are shown as a function of burst time-of-arrival (ToA) in Figure~\ref{figure: dm_toa}.  Clearly, there is a statistically significant variation in the observed DM among these bursts.  There are several ways to interpret this apparent DM variation. It could be that the actual DM (defined by Equation~\ref{equation: dm_define}) is changing. This could be the result of changing electron densities anywhere along the line of sight (LOS), including the environment near the source, the host galaxy, the intergalactic medium, or the contribution from the Milky Way.  Even short time scale variations could arise from motion through a dense environment near the source \citep{2020ApJ...893L..31Y}. However, it is also possible that the apparent change in DM is actually caused by confounding effects like intrinsic frequency drifting in the burst emission, frequency dependent burst morphology, and other propagation effects.

\begin{table}
\centering
\begin{tabular}{c|c}
\hline \hline
Burst-number & Optimized DM (pc cm$^{-3}$)  \\
\hline
02	&349.4(4)\\
03	&350.0(1)\\
04	&349.074(6)\\
05	&349.2(2)\\
06	&349.7(1)\\
07	&349.51(5)\\
11	&348.765(2)\\
12	&349.8(1)\\
13	&349.04(2)\\
14	&348.8(3)\\
15	&348.82(3)\\
\hline
\end{tabular}
\caption{\textbf{The optimized DM and the uncertainty.} We measure the optimized DM of eleven bursts in \citet{2020MNRAS.499L..16M} with the technique in this paper. The uncertainty corresponds to a 68$\%$ confidence interval.
}
\label{table: table_DM}
\end{table}

\subsection{The Effect of Noise on DM-power}\label{subsec: noise amplification}

\begin{figure*}
\centering     
\subfigure[p-28]{\label{fig:adding_noise_dmopt_p28}\includegraphics[width=0.32\textwidth]{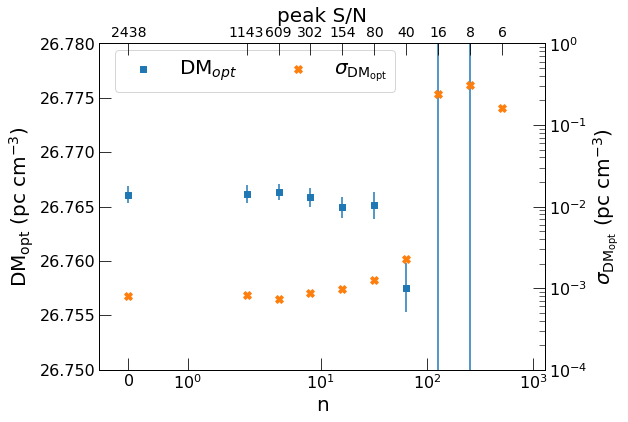}}
\subfigure[b-11]{\label{fig:adding_noise_dmopt_b11}\includegraphics[width=0.32\textwidth]{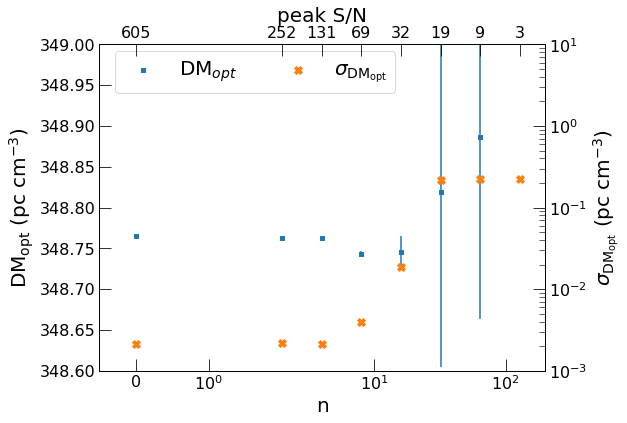}}
\subfigure[b-4]{\label{fig:adding_noise_dmopt_b4}\includegraphics[width=0.32\textwidth]{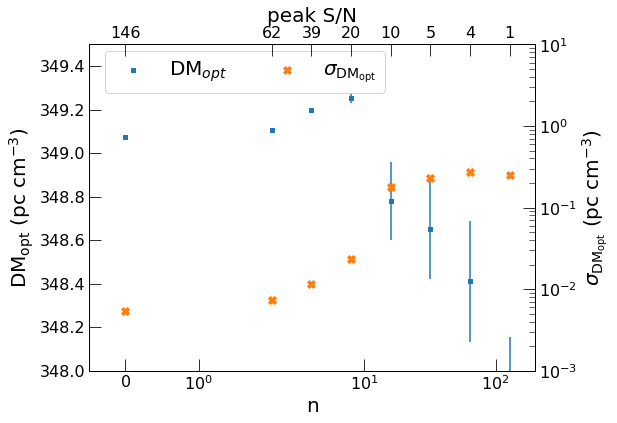}}
\caption{\textbf{The optimized DM and the uncertainty with additional noise.} The blue rectangle and the orange cross, corresponding to the left and right vertical axis, represent the optimized DM and the 68$\%$ confidence level at each of the noise factor \textit{n}. The \textit{n} is an integer that we amplify the standard deviation of each frequency channel of the off-pulse region with. We generate a Gaussian noise background with the same mean and amplified standard deviation, and add the background to the original dynamic spectrum. The corresponding peak S/N is in the top axis.}
\label{figure: adding_noise_dmopt_results}
\end{figure*}

To understand the interplay between the optimization technique and the noise background, we add noise to bursts b-11 and b-4 of FRB 20180916b as well as the 28th single pulse of PSR B0329+54 (hereafter: p-28). We measure the mean and standard deviation for each of the frequency channels in the off-pulse region in the dynamic spectrum. We amplify the standard deviation with a noise factor \textit{n}, generated a Gaussian noise background with the same mean and amplified standard deviation, and add the background to the original dynamic spectrum. Figures \ref{figure: adding_noise_dmopt_b11}, \ref{figure: adding_noise_dmopt_b4}, and \ref{figure: adding_noise_dmopt_p28} (See Appendix \ref{sec:The dynamic spectrum with adding noise}.) show the dynamic spectrum and the time-series with the various noise factors as well as the corresponding peak S/N. Figure \ref{figure: adding_noise_dmopt_results} shows the DM and uncertainty for the bursts with the additional noise. When the noise level is higher, the DM uncertainty is also going higher. Once the spectrum is dominated by the noise, the DM uncertainty would be large, which is consistent with what we expected.

\subsection{The Effect of Time Resolution}
Since FRBs can have complicated structure on a range of timescales, it may be possible to achieve more precise DM measurements by using higher time resolutions and thus sampling higher Fourier frequencies.  Here we consider the effect of time resolution on the DM measurement of two bursts from FRB~20180916B with very different morphologies.

Figure~\ref{figure: waterfalls_before_after} shows time-frequency waterfall plots and dedispersed profiles for bursts b-4 and b-11.  Burst b-11 has a complex structure with two bright narrow leading peaks and a broader trailing peak showing downward drifting behavior in frequency.  Burst b-4, on the other hand, only shows a fairly wide blob with downward drifting frequency structure. Figure \ref{figure: dm_opt_results} shows the $\rm DM_{\rm m}$ and $\sigma_{\rm DM_{\rm m}}$ measured in each of the log-Doppler frequency bins for both b-11 and b-4. For burst b-11, the $\sigma_{\rm DM_{\rm m}}$ decreases with increasing Doppler frequency, which suggests that higher time resolution would allow for a more precise measurement of the DM. For burst b-4, the behavior is quite different. The $\sigma_{\rm DM_{\rm m}}$ decreases for the first few Doppler frequency bins, but beyond a few kHz, it begins to rapidly increase again.  This means that the smallest time structures 
in burst b-4 are being resolved at this time resolution and higher time resolutions would not result in more precise DM measurements.  However, because we weight each $\rm DM_{\rm m}$ by $\sigma_{\rm DM_{\rm m}}$ (Section~\ref{subsec: Measure the FFT-mode independent DM and error}), the highly uncertain DM values measured at these Doppler frequencies will not affect the combined optimized DM for the burst.

\section{Summary}\label{sec: summary}

We have presented DM-power, a new algorithm to measure DM by weighting different Fourier modes in the power spectrum. This technique is motivated by the complex structure over a range of timescales seen in FRBs. We expect the DM-power technique to be particularly effective in measuring the DM of bursts with complex morphology (i.e. multiple components).

The DM-power method generates power spectrum from the de-dispersed profiles over a range of trial DMs and measures the Fourier power in several logarithmically spaced Doppler frequency bins.  By fitting Gaussian functions to the DM power, the DM and DM uncertainty can be measured in each Doppler frequency bin. The measurements from each Doppler frequency bin are weighted by the variance and combined to make a global DM and DM uncertainty for the burst. These procedures have been validated using 56 single pulses from the pulsar PSR B0329+54, which has a daily DM variation of less than  10$^{-4}$ pc cm$^{-3}$.

We applied our DM-power technique to 11 bright FRB~20180916B bursts from \citet{2020MNRAS.499L..16M} and find that the bursts display a statistically significant variation in observed DM over the course of about 2~hours. While this may be the result of a change in the integrated electron density along the line of sight (potentially in a high density environment around the source), the fact that the changes are not smooth with time suggests that this is not the case.  Instead, we find it more likely that the observed variation in DM is actually caused by other confounding effects like intrinsic frequency dependent burst structure and possibly other propagation effects. Another reason for the DM variation is that the DM uncertainties are underestimated due to RFI contamination, which we encourage future study on the interplay between DM uncertainty and RFI contamination.

The DM-power technique allows for precise measurements of burst DMs. The ultimate precision of this method will of course depend on various observational parameters and burst properties, but for the sources considered in this paper we were able to get DM measurements down to $\sim$ 0.001~pc~cm $^{-3}$. We encourage the community to measure the DM variation of repeaters by using DM-power. If the DM variation is measured, it would be a hint to understand the local environment of repeaters.

\section*{Acknowledgements}
We thank Marten van Kerkwijk for useful discussions on relevant statistical techniques. We thank all the members of the Scintillometry group at CITA at the University of Toronto for useful discussions. We thank the staff of the GMRT that made these observations possible. GMRT is run by the National Centre for Radio Astrophysics of the Tata Institute of Fundamental Research. VRM acknowledges support of the Department of Atomic Energy, Government of India, under project no. 12-R$\&$D-TFR-5.02-0700. Ue-Li Pen receives support from Ontario Research Fund—research Excellence Program (ORF-RE), Natural Sciences and Engineering Research Council of Canada (NSERC) [funding reference number RGPIN-2019-067, CRD 523638-18, 555585-20], Canadian Institute for Advanced Research (CIFAR), the National Science Foundation of China (Grants No. 11929301), Thoth Technology Inc, Alexander von Humboldt Foundation, and the National Science and Technology Council (NSTC) of Taiwan (111-2123-M-001 -008 -, and 111-2811-M-001 -040 -). Computations were performed on the SOSCIP Consortium’s [Blue Gene/Q, Cloud Data Analytics, Agile and/or Large Memory System] computing platform(s). SOSCIP is funded by the Federal Economic Development Agency of Southern Ontario, the Province of Ontario, IBM Canada Ltd., Ontario Centres of Excellence, Mitacs and 15 Ontario academic member institutions. LGS is a Lise Meitner Max Planck Group leader and acknowledges funding from the Max Planck Society. R.S.W. is supported by an appointment to the NASA Postdoctoral Program at the Jet Propulsion Laboratory, administered by Oak Ridge Associated Universities under contract with NASA. Part of this research was carried out at the Jet Propulsion Laboratory, California Institute of Technology, under a contract with the National Aeronautics and Space Administration. Computations were performed on the Niagara and Cedar supercomputers at the SciNet HPC Consortium \citep{2010JPhCS.256a2026L,2019arXiv190713600P} as well as the CITA cluster. SciNet is funded by: the Canada Foundation for Innovation; the Government of Ontario; the Ontario Research Fund - Research Excellence; and the University of Toronto. 

\section*{Data Availability}
The data analyzed in this study will be shared on reasonable request to the corresponding author. The DM-power package is available on {the GitHub page}{https://github.com/hsiuhsil/DM-power}.

\bibliographystyle{mnras}
\bibliography{main} 

\appendix

\section{The simulated Gaussian pulse profile}\label{sec:The simulated Gaussian pulse profile}

In this section, we show the simulated Gaussian profiles as mentioned in Section \ref{subsub: the simulated Gaussian pulse profiles}.

\begin{figure}
\includegraphics[width=0.4\textwidth]{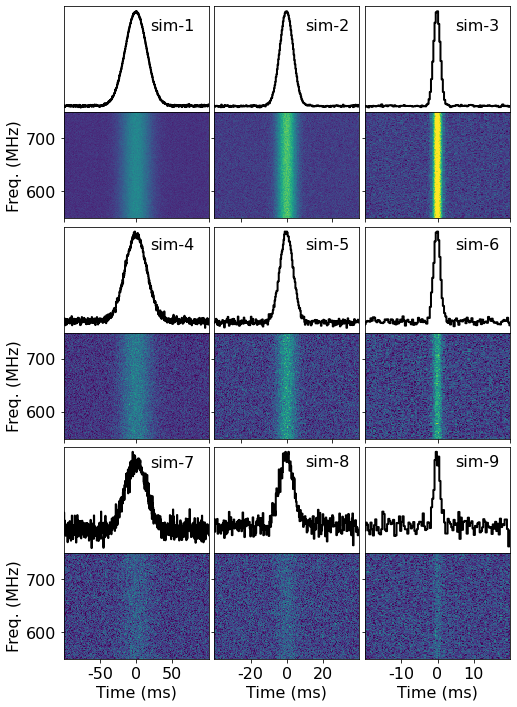}
\caption{\textbf{The 9 simulated Gaussian pulse profiles.} In each panel, the lower and upper panel shows the dynamic spectrum and the time-series data (i.e. the average of the dynamic spectrum in the frequency axis.), respectively. The peak S/N of the first, second, and the third row is 200, 50, and 15, respectively. The width of the first, second, and the third column is 33.2, 8.3, and 1.7 ms, respectively.
}
\label{figure: gaussian_sim_pulses}
\end{figure}

\section{The dynamic spectrum before and after the DM optimization}\label{sec:The dynamic spectrum before and after the DM optimization}

We demonstrate the dynamic spectrum of bursts from FRB 20180916B, as mentioned in Section \ref{sub:FRB 20180916B}, with the DM before and after the optimization procedures.

\begin{figure*}
\includegraphics[width=0.9\textwidth]{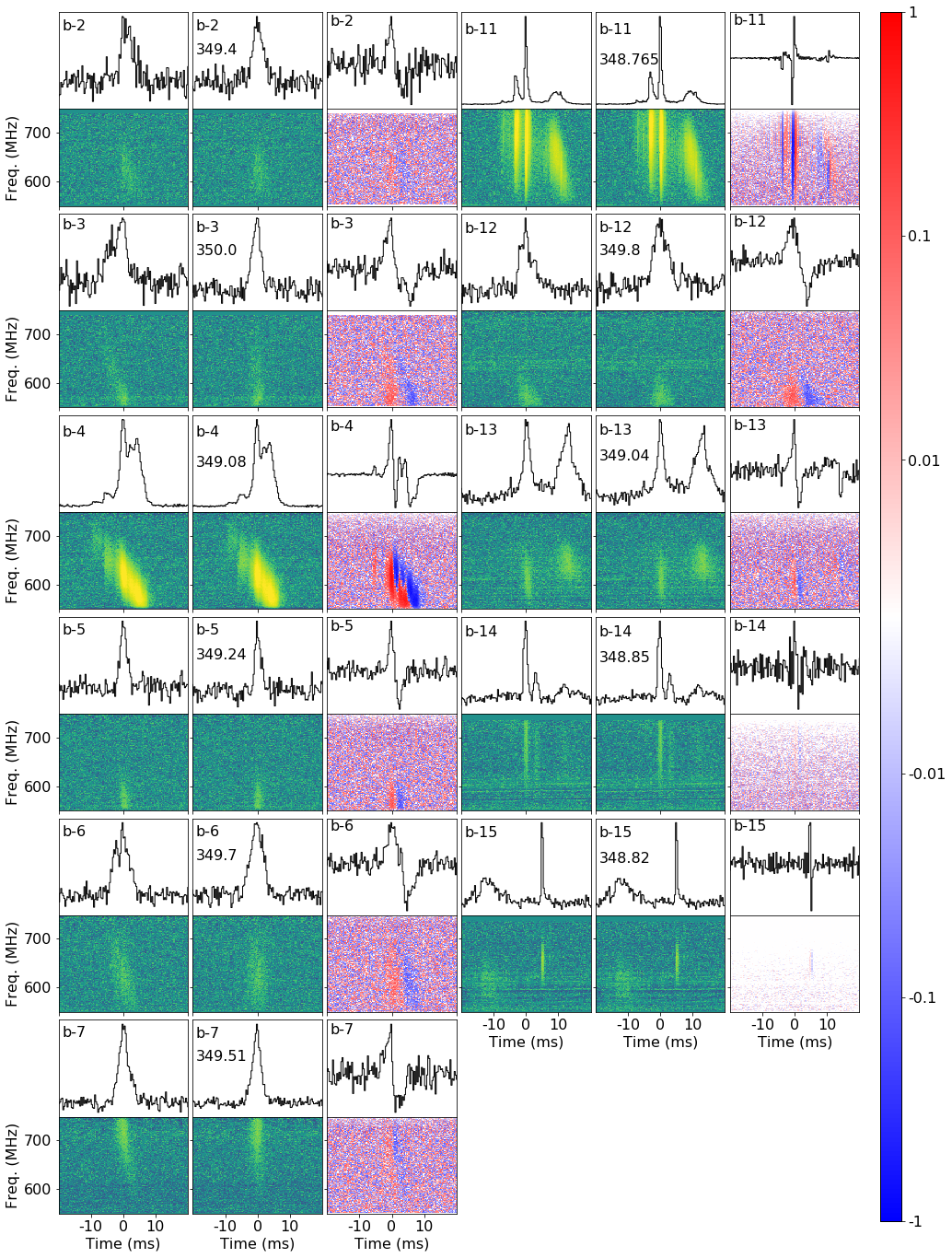}
\caption{\textbf{The waterfall with the original and optimized DM as well as their difference.} The first and fourth columns show the waterfall of the 11 bursts at the original DM (348.82 pc cm$^{-3}$). The second and the fifth columns show the waterfall of the 11 bursts at the correspondingly optimized DM. The third and the sixth columns show the difference between the optimized and the original case. 
}
\label{figure: waterfalls_res}
\end{figure*}

\section{The dynamic spectrum of single pulses from PSR B0329+54}\label{sec:The dynamic spectrum from PSR B0329+54}

We show the dynamic spectrum of giant pulses from PSR B0329+54, as mentioned in Section \ref{subsec: 56 pulses from PSR B0329+54}, with the DM before and after the optimization procedures.

\begin{figure*}
\includegraphics[width=0.85\textwidth]{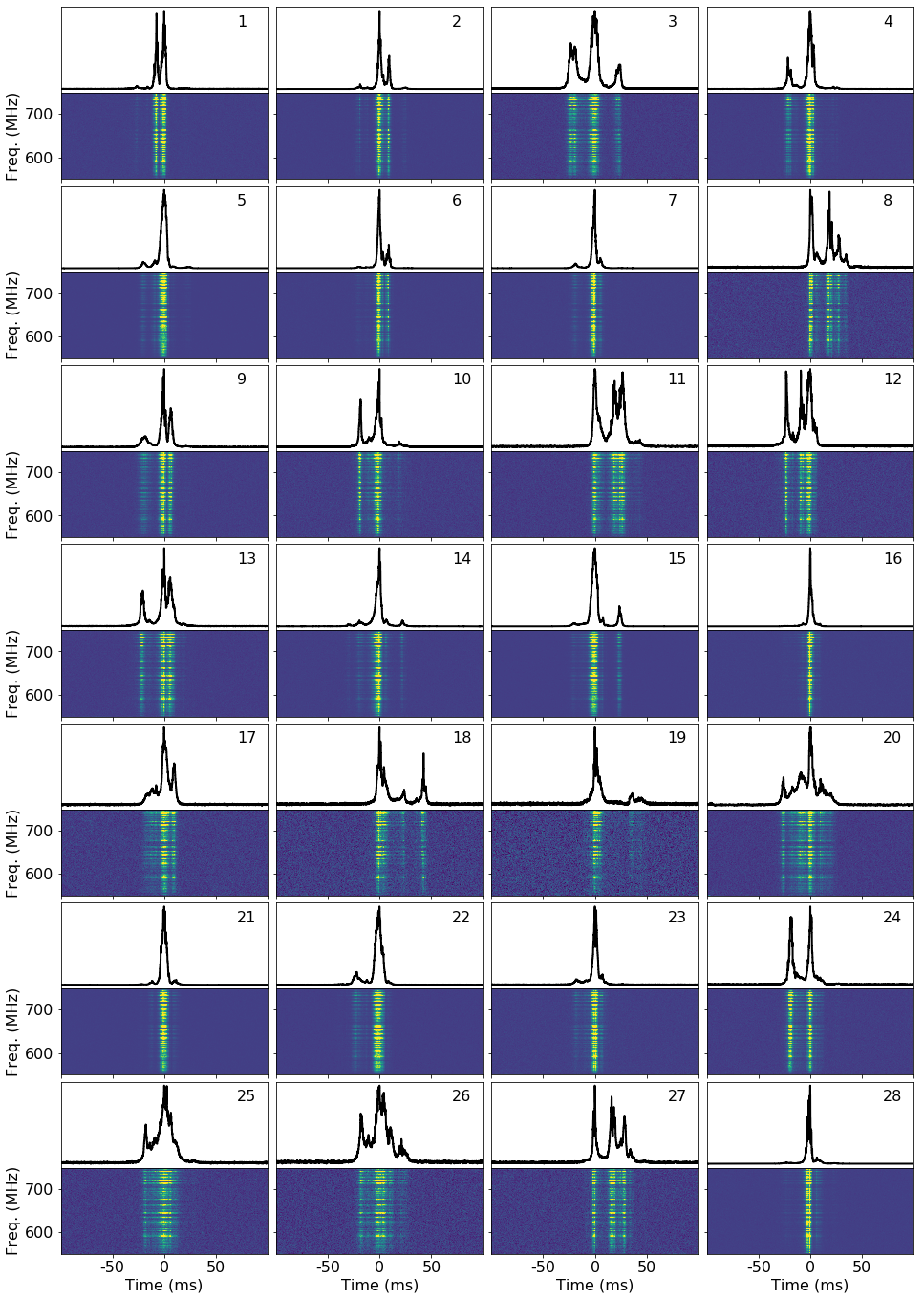}
\caption{\textbf{The dynamic spectrum of single pulses 1-28 from PSR B0329+54.} Each pulse was de-dispersed with the optimized DM value. In each panel, the bottom shows the dynamic spectrum with the optimized DM, where the time resolution is 81.92 $\mu$s and the frequency resolution is rebinned to 1.5625 kHz, respectively, and the top shows the corresponding time-series data. }
\label{figure: b0329_gallery_1}
\end{figure*}

\begin{figure*}
\includegraphics[width=0.85\textwidth]{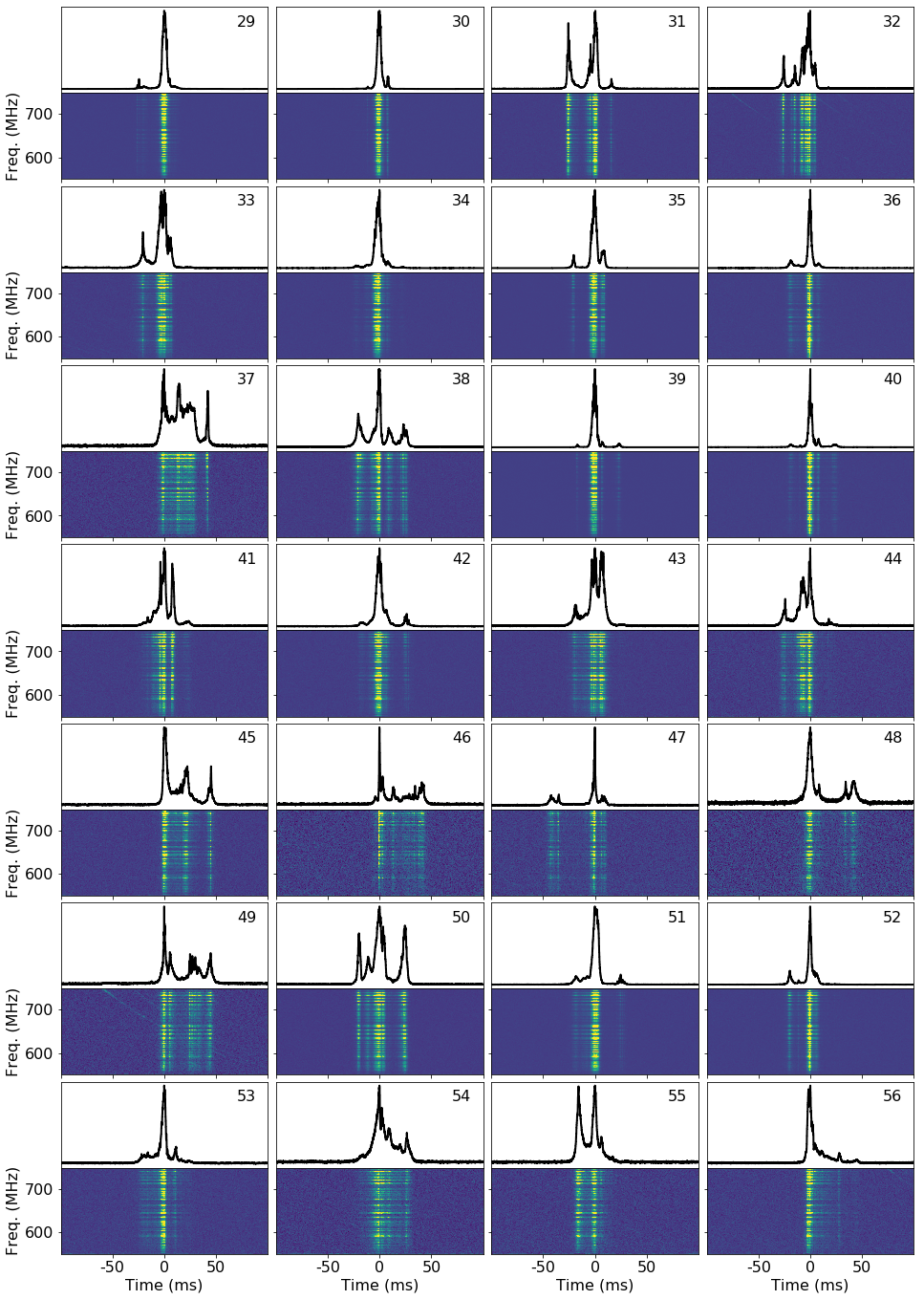}
\caption{\textbf{The dynamic spectrum of single pulses 29-56 from PSR B0329+54.} The same description as mentioned in Figure \ref{figure: b0329_gallery_1}.}
\label{figure: b0329_gallery_2}
\end{figure*}

\section{The dynamic spectrum with adding noise}\label{sec:The dynamic spectrum with adding noise}

In this section, we show the dynamic spectrum after adding the noise, which we mentioned the procedures in detail in Section \ref{subsec: noise amplification}.

\begin{figure*}
\includegraphics[width=0.9\textwidth]{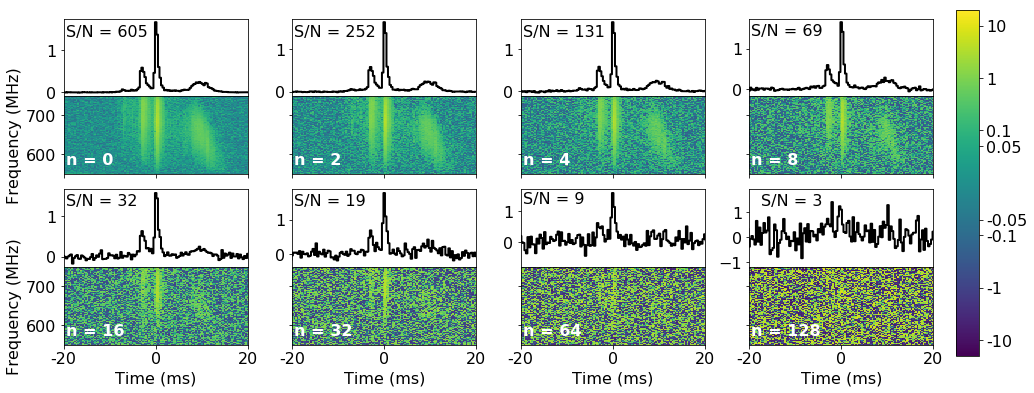}
\caption{\textbf{The dynamic spectrum with adding noise and the time series of b-11.} In each panel, the bottom shows the dynamic spectrum with the adding noise, where the noise factor and the corresponding time-series are shown in the top
}
\label{figure: adding_noise_dmopt_b11}
\end{figure*}

\begin{figure*}
\includegraphics[width=0.9\textwidth]{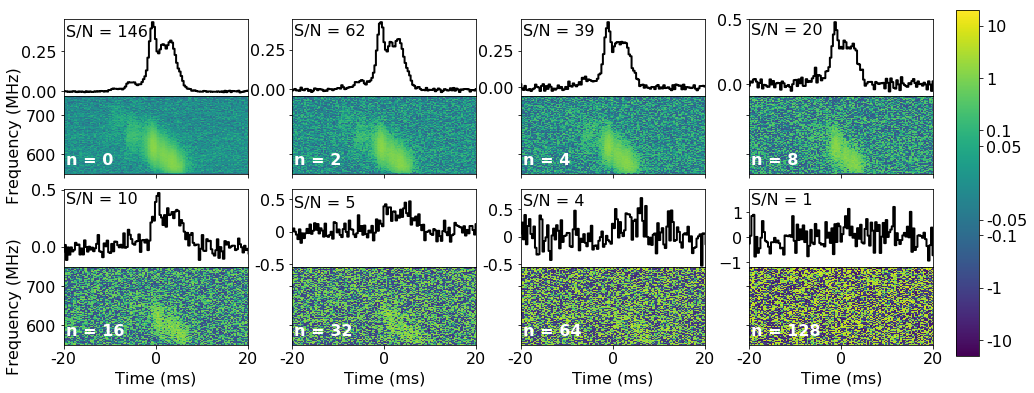}
\caption{\textbf{The dynamic spectrum with adding noise and the time series of b-4.} The same description as mentioned in Figure \ref{figure: adding_noise_dmopt_b11}.}
\label{figure: adding_noise_dmopt_b4}
\end{figure*}

\begin{figure*}
\includegraphics[width=0.9\textwidth]{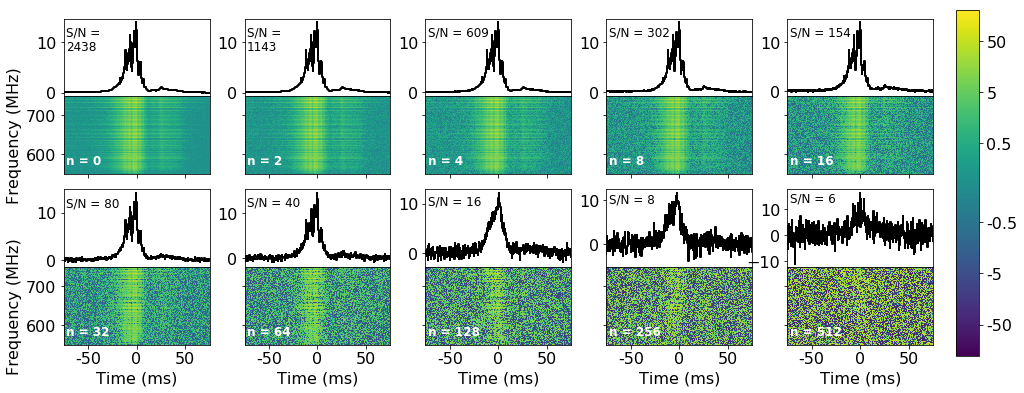}
\caption{\textbf{The dynamic spectrum with adding noise and the time series of p-28 of B0329+54.} The same description as mentioned in Figure \ref{figure: adding_noise_dmopt_b11}.}
\label{figure: adding_noise_dmopt_p28}
\end{figure*}


\bsp	
\label{lastpage}
\end{document}